\documentclass[a4paper,11pt]{article}
\usepackage{amsmath}
\usepackage{graphicx}
\usepackage{amssymb}
\usepackage{natbib}
\usepackage[hidelinks]{hyperref}

\hypersetup{
  colorlinks = true,     
  urlcolor = blue,       
  linkcolor = blue,      
  citecolor = red         
}

\renewcommand{\mod}[1]{\left|#1\right|}
\renewcommand{\vec}[1]{\mathbf{#1}}
\renewcommand{\S}{\mathcal{S}}
\renewcommand{\H}{\mathcal{H}}

\newcommand{\sqmod}[1]{\left|#1\right|^{2}}
\newcommand{\expt}[1]{\left<#1\right>}

\newcommand{\rmi}{\mathrm{i}}
\newcommand{\rme}{\mathrm{e}}
\newcommand{\E}{\epsilon}
\newcommand{\f}{\phi}
\newcommand{\R}{\mathcal{R}}
\newcommand{\Env}{\mathcal{E}}
\newcommand{\w}{\omega}

\newcommand{\bra}[1]{\left<#1\right|}
\newcommand{\ket}[1]{\left|#1\right>}

\newcommand{\iprod}[2]{\left<#1|#2\right>}
\newcommand{\iprodM}[3]{\left<#1\right|#2\left|#3\right>}

\newcommand{\oprod}[2]{\left|#1\right>\left<#2\right|}

\providecommand{\keywords}[1]
{
  \small	
  \textbf{\textit{Keywords---}} #1
}

\begin{document}

\title{Physical Observables and Superselection}

\author{M.P. Vaughan\footnote{\emph{Corresponding author}: martin.vaughan@physics.org}}

\date{}

\maketitle

\begin{abstract}
It is well-known that a physical state may be represented by a unit vector in a Hilbert space, whilst a physical observable may be represented by a self-adjoint operator on that same space. However, the introduction of a superselection rule on a Hilbert space imposes constraints such that not all self-adjoint operators can represent observables, which has further consequences for the states themselves as well as the physical processes that transform them. In this article, we review the concept of superselection from the view of how such rules would affect what we consider to be physical. Arguments for the existence of a superselection rule due to charge or possibly energy conservation are given and the emergence of effective superselection rules due to decoherence or the lack of a shared reference frame are also discussed.  
\end{abstract}


\tableofcontents

\section{Introduction}
Since von Neumann's recognition that a quantum state can be represented by a unit vector of a Hilbert space~\cite{vonNeumann1932mathematical}, it has been taken as a postulate of quantum mechanics that a \emph{physical state} can be represented mathematically by such a state vector. Previously, Dirac had actually gone beyond this~\cite{dirac1981principles} and introduced a formalism that relied on a more general structure than a Hilbert space (where a Hilbert space can only have either finite or countably infinite dimensions). However, we shall not be venturing beyond the Hilbert space formulation in this article but shall confine ourselves to what we may call the `conventional' postulates of quantum mechanics, which still cover much of Dirac's work.

A second postulate is that \emph{physical observables} (the quantities that experimentalists actually measure) are represented by self-adjoint operators on the same Hilbert space. Schr{\"o}dinger's equation may then be incorporated by a further postulate that the time evolution of a state is described by a unitary transformation formulated in terms of the Hamiltonian of the system. 

Questions that may then occur to the student of quantum mechanics on first becoming acquainted with these postulates are: \emph{does \emph{every} unit vector of a Hilbert space represent a valid physical state} and \emph{does \emph{every} self-adjoint operator represent a valid physical observable?} One might argue that there is no \emph{necessary} reason for this to be so. On the other hand, if this is not the case, how should we distinguish those states or observables that are `physical' from those that are not?

This is a deep philosophical question that touches on the question of what we mean when we say that something is `physical'. As physicists, though, the question is of more immediate practical importance. We need to have a valid mathematics to formulate the physics but we also need to know whether the mathematics is \emph{always} describing valid physics.

These questions become particularly pertinent when we come to consider the possibility of \emph{superselection rules}. These rules (appear to) come about via the imposition of certain physical considerations (although it would be important to confirm that they do not emerge from any mathematical necessity). However, if such rules do exist, then they also have the effect of constraining what we consider to be the \emph{physical observables} of a system. We shall also see that this constrains what can be a physical \emph{states} of a system as well as what constitutes a valid physical \emph{process}. Given the very powerful arguments for the existence of such rules, it will be of benefit to have a good understanding of them and their implications.  

\section{Superselection rules}
Superselection is a concept in quantum mechanics that implies that, for certain physical observables, coherent superpositions of eigenstates with different eigenvalues are not possible. This involves the idea that the Hilbert space of the system can be decomposed into irreducible components (known as superselection sectors) and that no operator representing a physically realisable observable can couple states from different sectors. An interpretation of this is that a superselection rule imposes restrictions on exactly what can be a physical observable of the system.

\subsection{Definition}\label{sec:definition}
An ordinary selection rule between two states $\ket{\psi_{1}}$ and $\ket{\psi_{2}}$ is said to pertain if

\begin{align}
\iprodM{\psi_{1}}{\hat{H}}{\psi_{2}} &= 0, \nonumber
\end{align}

\noindent where $\hat{H}$ is the Hamiltonian of the system. If for \emph{any} physically realisable observable $\hat{A}$, we have 

\begin{align}
\iprodM{\psi_{1}}{\hat{A}}{\psi_{2}} &= 0, \label{eq:superselection-rule}
\end{align}

\noindent then $\ket{\psi_{1}}$ and $\ket{\psi_{2}}$ are said to be separated by a \emph{superselection rule}. 

The very definition of such a rule implies that not every self-adjoint operator represents a physically realisable observable. For example, for any two states $\ket{\psi_{1}}$ and $\ket{\psi_{2}}$ that are claimed to be separated by a superselection rule, we may always construct the operator

\begin{align}
\hat{B} &= \oprod{\psi_{1}}{\psi_{2}} + \oprod{\psi_{2}}{\psi_{1}}. \nonumber
\end{align}

\noindent This is evidently self-adjoint and yet

\begin{align}
\iprodM{\psi_{1}}{\hat{B}}{\psi_{2}} &= 1. \nonumber
\end{align}

If a superselection does truly pertain, then we must insist that $\hat{B}$ \emph{cannot} represent a physical observable. This clearly begs the question: \emph{when is a self-adjoint operator not a physical observable}? The answer to this turns out to be somewhat circular, since, as we shall argue in Section~\ref{eq:observables}, the example of $\hat{B}$ cannot be an observable \emph{because} $\ket{\psi_{1}}$ and $\ket{\psi_{2}}$ are separated by a superselection rule. It may then be argued that the \emph{assertion} of a superselection rule \emph{constrains} what the valid physical observables of the theory can be. 

\subsection{Superselection sectors}\label{sec:direct-sum}
Any Hilbert space $\H$ may be decomposed into a direct sum of orthogonal subspaces, $\H_{n}$, via

\begin{align}
\H &= \H_{1} \oplus \H_{2} \oplus \ldots \oplus \H_{N},  \nonumber \\
&= \bigoplus_{n=1}^{N} \H_{n} \label{eq:direct-sum}
\end{align}

\noindent Superselection rules as defined by \eqref{eq:superselection-rule} will arise if these subspaces constitute what are called \emph{superselection sectors}. This will be the case if every physical observable can be represented by a direct sum of operators, each acting on just one subspace $\H_{n}$. That is, every observable must be represented in the block diagonal form

\begin{align}
\hat{A} &= \left[\begin{array}{cccc}
\hat{A}_{1} & \vec{0} & \ldots & \vec{0} \\
\vec{0} & \hat{A}_{2} & \ldots & \vec{0} \\
\vdots & \vdots & \ddots & \vdots \\
\vec{0} &  \vec{0} &  \ldots & \hat{A}_{N}\end{array}\right].  \nonumber
\end{align}

\noindent where each sub-matrix $\hat{A}_{n}$ acts on just one subspace $\H_{n}$.

As a simple example, suppose $\H$ may be decomposed into exactly 2 superselection sectors $\H_{1}$ and $\H_{2}$ so that every physical observable has the block-diagonal matrix form

\begin{align}
\hat{A} &= \left[\begin{array}{cc}
\hat{A}_{1} & \vec{0} \\
\vec{0} & \hat{A}_{2}\end{array}\right].  \nonumber
\end{align}

\noindent A general ket of the Hilbert space may then be written similarly in block form as

\begin{align}
\ket{\psi} &= \left[\begin{array}{c}
\ket{\phi_{1}} \\
\ket{\phi_{2}} \end{array}\right].  \nonumber
\end{align}

\noindent Vectors lying solely on either $\H_{1}$ or $\H_{2}$ will then have the form

\begin{align}
\ket{\psi_{1}} &= \left[\begin{array}{c}
\ket{\phi_{1}} \\
\vec{0} \end{array}\right]~\mathrm{and}~\ket{\psi_{2}} = \left[\begin{array}{c}
\vec{0} \\
\ket{\phi_{2}} \end{array}\right] \nonumber
\end{align}

\noindent respectively. It is then clear to see that

\begin{align}
\left[\begin{array}{cc}
\bra{\phi_{1}} & \vec{0}\end{array}\right]\left[\begin{array}{cc}
\hat{A}_{1} & \vec{0} \\
\vec{0} & \hat{A}_{2}\end{array}\right]\left[\begin{array}{c}
\vec{0} \\
\ket{\phi_{2}} \end{array}\right] &= 0.  \nonumber
\end{align}

We may represent the general case more concisely by defining $\Pi_{n}$ to be the projector onto the subspace $\H_{n}$. That is,

\begin{align}
\Pi_{n} &\equiv \sum_{\ket{\phi_{i}}\in\H_{n}} \ket{\phi_{i}}\bra{\phi_{i}}. \label{eq:Pi-def}
\end{align}

\noindent A block diagonal operator may now be written explicitly as

\begin{align}
\hat{A} &= \sum_{n}^{N}\Pi_{n}\hat{A}_{n}\Pi_{n}.  \label{eq:hat-A-block}
\end{align}

\noindent This form makes in clear that the operator cannot couple states from different $\H_{n}$.

\subsection{Superselection and charge conservation}\label{sec:charge}
When Wick \emph{et al} first introduced the concept of superselection, they offered a plausibility argument that such a rule should exist between eigenstates of the total charge operator~\cite{wick1952intrinsic}. This may be taken as a consequence of the \emph{assertion} that the total charge operator commutes with all other observables.

Let us consider a general operator $\hat{Q}$ (that we might possibly take to be the total charge operator) and assume that it commutes with all other observable operators on a given Hilbert space $\H$. That is,

\begin{align}
[\hat{Q}, \hat{A}] &= \hat{Q}\hat{A} - \hat{A}\hat{Q} = 0, \nonumber
\end{align}

\noindent for \emph{all} observables $\hat{A}$ on $\H$. This then has the consequence that no observable $\hat{A}$ can change the `charge' $q$ of a state initially in an eigenstate of $\hat{Q}$ with eigenvalue $q$.

Specifically, let the eigenvalues and eigenvectors of $\hat{Q}$ be $q$ and $\ket{\phi_{q}}$ respectively. Then, since every observable commutes with $\hat{Q}$, we have

\begin{align}
\hat{Q}\hat{A}\ket{\phi_{q}} &= \hat{A}\hat{Q}\ket{\phi_{q}} = q\hat{A}\ket{\phi_{q}}. \nonumber
\end{align}

\noindent Now, in general, $\hat{A}\ket{\phi_{q}}$ may be a different state to $\ket{\phi_{q}}$ (as well as possibly being multiplied by a scalar value) but it must still be proportional to an eigenvector of $\hat{Q}$ with the same eigenvalue $q$. Hence, no observable operator can change the value of $q$ (which is certainly a plausible argument in the case of electric charge). Moreover, since any eigenvector $\ket{\phi_{q'}}$ with a \emph{different} eigenvalue $q' \ne q$ must be orthogonal to both $\ket{\phi_{q}}$ and $\hat{A}\ket{\phi_{q}}$, we will also have

\begin{align}
\iprodM{\phi_{q'}}{\hat{A}}{\phi_{q}} &= 0,~(q'\neq q), \label{eq:charge-super}
\end{align}

\noindent which is just the criterion of \eqref{eq:superselection-rule} for a superselection rule.

We may now construct the direct sum of Hilbert spaces (as specified in \eqref{eq:direct-sum}) by defining $\H_{q}$ to be the subspace of $\H$ spanned by the eigenvectors of $\hat{Q}$ with eigenvalue $q$. One can then write

\begin{align}
\H &=  \bigoplus_{q} \H_{q}, \nonumber
\end{align}

\noindent where the sum is over all possible values of $q$. Since no observable operator can take a vector from $\H_{q}$ and transform it into a different subspace $\H_{q'}$, all such operators must necessarily take the block diagonal form described in Section~\ref{sec:direct-sum} above.

So far we have seen that a superselection rule exists on a Hilbert space if a particular observable $\hat{Q}$ commutes with all other observables. We may also prove the converse. That is, if a superselection rule pertains between those eigenvectors of an operator $\hat{Q}$ having distinct eigenvalues, then $\hat{Q}$ commutes with all observables. 

We do this by constructing two general state vectors $\ket{\psi}$ and $\ket{\psi'}$ from the eigenvectors of $\hat{Q}$ so that we have

\begin{align}
\ket{\psi} = \sum_{q} a_{q}\ket{\phi_{q}}~\mathrm{and}~\ket{\psi'} = \sum_{q} b_{q}\ket{\phi_{q}}, \nonumber
\end{align}

\noindent where $a_{q} = \iprod{\psi}{\phi_{q}}$ and $b_{q} = \iprod{\psi'}{\phi_{q}}$. We then consider the inner product involving an arbitrary observable $\hat{A}$

\begin{align}
\iprodM{\psi'}{\hat{A}\hat{Q}}{\psi} &= \sum_{q'q} qa_{q}b_{q'}^{*}\iprodM{\phi_{q'}}{\hat{A}}{\phi_{q}}. \nonumber
\end{align}

\noindent Now, by assumption, a superselection rule exists between any states $\ket{\phi_{q'}}$ and $\ket{\phi_{q}}$ where $q' \ne q$, so the double sum reduces to a single sum over $q$

\begin{align}
\iprodM{\psi'}{\hat{A}\hat{Q}}{\psi} = \sum_{q} qa_{q}b_{q}^{*}\iprodM{\phi_{q}}{\hat{A}}{\phi_{q}}. \nonumber
\end{align}

Next, we consider the product $\iprodM{\psi'}{\hat{Q}\hat{A}}{\psi}$, where we take $\hat{Q}$ to be acting on $\bra{\psi'}$ to the right to give

\begin{align}
\iprodM{\psi'}{\hat{Q}\hat{A}}{\psi} &= \sum_{q'q} q'a_{q}b_{q'}^{*}\iprodM{\phi_{q'}}{\hat{A}}{\phi_{q}}, \nonumber \\
&= \sum_{q} qa_{q}b_{q}^{*}\iprodM{\phi_{q}}{\hat{A}}{\phi_{q}}. \nonumber
\end{align}

\noindent This is just the same result as before and we evidently have

\begin{align}
\iprodM{\psi'}{\hat{Q}\hat{A}}{\psi} - \iprodM{\psi'}{\hat{A}\hat{Q}}{\psi} &= \iprodM{\psi'}{[\hat{Q},\hat{A}]}{\psi} = 0. \nonumber
\end{align}

\noindent Since $\ket{\psi}$ and $\ket{\psi'}$ are arbitrary, we therefore conclude

\begin{align}
[\hat{Q},\hat{A}] &= 0. \nonumber
\end{align}

\noindent That is, $\hat{Q}$ commutes with any physical observable $\hat{A}$. In conclusion, then, a superselection rule will exist on a Hilbert space \emph{if and only if} there exists a operator  $\hat{Q}$ representing some physical observable that commutes with all other physical observables.

It has since been shown by Strocchi and Wightman~\cite{strocchi1974proof} in relativistic quantum field theory that the total charge operator $\hat{Q}$ does indeed commute with all \emph{quasilocal} operators, which therefore implies the existence of a superselection rule. Here, the adjective `quasilocal' means that an operator depends only on the value of fields in a local region $\Omega$ around a given spacetime coordinate. Whilst Strocchi and Wightman's actual proof is rather long and involved, they do give a much simpler, intuitive argument based on Gauss' law of electrostatics.

Consider a general, quasilocal operator $\hat{A}$ representing a physical observable on a bounded region of space $\Omega$. The total charge operator may be expressed in terms of charge density $\rho$ via

\begin{align}
\hat{Q} &= \lim_{R\to\infty}\int_{|\mathbf{r}| \le R} \rho(\mathbf{r})d^{3}\mathbf{r}, \nonumber
\end{align}

\noindent where, for simplicity, we can consider the volume integral to be over a sphere of radius $R$. The commutator of $\hat{Q}$ and $\hat{A}$ may then be written as

\begin{align}
[\hat{Q},\hat{A}] &= \lim_{R\to\infty}\int_{|\mathbf{r}| \le R} [\rho(\mathbf{r}), \hat{A}] d^{3}\mathbf{r}. \nonumber
\end{align}

Now, by Gauss' law, the charge density is related to the electric field $\mathbf{E}$ by

\begin{align}
\rho(\mathbf{r}) &= \epsilon_{0}\nabla\cdot\mathbf{E}(\mathbf{r}), \nonumber
\end{align}

\noindent where $\epsilon_{0}$ is the permittivity of free space. Using the divergence theorem, the commutator then becomes

\begin{align}
[\hat{Q},\hat{A}] &= \lim_{R\to\infty}\int_{|\mathbf{r}| = R} [\mathbf{E}(\mathbf{r}), \hat{A}]\cdot d\mathbf{S}, \nonumber
\end{align}

\noindent where the integral is now over the surface $S$ of the sphere. However, as we take the limit as $R\to\infty$, the surface over which we evaluate $\mathbf{E}$ will be outside the bounding region $\Omega$ on which $\hat{A}$ operates. Since $\hat{A}$ cannot depend on the field values in these regions of space, the commutator must vanish and we will have

\begin{align}
[\hat{Q},\hat{A}] &= 0 \nonumber
\end{align}

\noindent for all such quasilocal variables.

\subsection{Consequences for superpositions}\label{sec:superpositions}
The superselection rule of \eqref{eq:superselection-rule} will have consequences for the superposition of states that it separates. Ordinarily, we would construct a superposition by multiplying each of the component states by a complex amplitude and adding them together, giving, for example

\begin{align}
\ket{\psi} &= \alpha\ket{\psi_{1}} + \beta\ket{\psi_{2}}, \nonumber
\end{align}

\noindent normalised so that $\sqmod{\alpha} + \sqmod{\beta} = 1$. Since the overall phase of $\ket{\psi}$ has no observable consequence, we can write this as

\begin{align}
\ket{\psi} &= \mod{\alpha}\ket{\psi_{1}} + \mod{\beta}\rme^{\rmi\theta}\ket{\psi_{2}}, \label{eq:coherent}
\end{align}

\noindent where $\theta$ is the phase difference between the two components.

The expectation value of an observable $\hat{A}$ when the system is in this state is then calculated via

\begin{align}
\iprodM{\psi}{\hat{A}}{\psi} &= \sqmod{\alpha}\iprodM{\psi_{1}}{\hat{A}}{\psi_{1}} + \sqmod{\beta}\iprodM{\psi_{2}}{\hat{A}}{\psi_{2}} \nonumber \\
&+ 2\mod{\alpha}\mod{\beta}\mod{\iprodM{\psi_{1}}{\hat{A}}{\psi_{2}}}\cos\theta, \nonumber
\end{align}

\noindent (where any phase difference between $\iprodM{\psi_{1}}{\hat{A}}{\psi_{2}}$ and $\iprodM{\psi_{2}}{\hat{A}}{\psi_{1}}$ is assumed to be absorbed into $\theta$). The phase difference $\theta$ can then give rise to observable interference effects and \eqref{eq:coherent} is said to be in a \emph{coherent} superposition.

However, if a superselection rule between $\ket{\psi_{1}}$ and $\ket{\psi_{2}}$ exists then the interference term disappears and we are left with

\begin{align}
\iprodM{\psi}{\hat{A}}{\psi} &= \sqmod{\alpha}\iprodM{\psi_{1}}{\hat{A}}{\psi_{1}} + \sqmod{\beta}\iprodM{\psi_{2}}{\hat{A}}{\psi_{2}}. \nonumber
\end{align}

\noindent This is the result for a \emph{classical} mixture of the two states, in which we have no information about any phase relation between $\ket{\psi_{1}}$ and $\ket{\psi_{2}}$. It is therefore not possible to have a \emph{coherent} superposition of the two states. That is, since no physical observation can give us any information about $\theta$, the phase difference is unobservable. Without a well-defined phase difference, we do not have a coherent superposition. An expression such as \eqref{eq:coherent} implying a phase difference between states from different superselection sectors would therefore be meaningless. Whilst we might write it down mathematically, such an expression would not be representing a \emph{physical} state that the system could actually be in.

This last point is perhaps made clearer by expressing \eqref{eq:coherent} as a density matrix. Mathematically, we could construct this as

\begin{align}
\ket{\psi}\bra{\psi} &= \sqmod{\alpha}\ket{\psi_{1}}\bra{\psi_1} + \sqmod{\beta}\ket{\psi_{2}}\bra{\psi_{2}} \nonumber \\
&+ \mod{\alpha}\mod{\beta}\rme^{-\rmi\theta}\ket{\psi_{1}}\bra{\psi_{2}} + \mod{\alpha}\mod{\beta}\rme^{\rmi\theta}\ket{\psi_{2}}\bra{\psi_1}. \label{eq:pure-state}
\end{align}

\noindent If the phase angle $\theta$ could actually be defined, this would represent a pure state density matrix. However, if the phase cannot be defined, then all we meaningfully have is

\begin{align}
\ket{\psi}\bra{\psi} &\to \sqmod{\alpha}\ket{\psi_{1}}\bra{\psi_1} + \sqmod{\beta}\ket{\psi_{2}}\bra{\psi_{2}}, \nonumber 
\end{align}

\noindent which is the density matrix of an ensemble of \emph{two} states, $\ket{\psi_{1}}$ and $\ket{\psi_{2}}$.

\subsection{Effective density matrix}
The loss of phase information discussed above in Section~\ref{sec:superpositions} may be formalised in terms of the density matrix of the system $\hat{\rho} = \oprod{\psi}{\psi}$. In the density matrix, relative phase information is encoded into the off-diagonal elements, which we refer to as \emph{coherences}. We shall now see how the assumption of a superselection rule is equivalent to the assumption of an effective density matrix in which the coherences between different subspaces have been removed.

The general expression for the expectation value of an operator $\hat{A}$ in terms of the density matrix is

\begin{align}
\langle\hat{A}\rangle &= \mathrm{Tr}[\hat{\rho}\hat{A}], \label{eq:expt-trace}
\end{align}

\noindent where $\mathrm{Tr}[\hat{M}]$ denotes the trace of a matrix $\hat{M}$. Let us suppose that the Hilbert space of the system can be decomposed into a direct sum of subspaces $\H_{n}$ as in \eqref{eq:direct-sum} and that $\hat{A}$ is a block diagonal operator as defined by \eqref{eq:hat-A-block}. 

Since the sum of the projectors over all subspaces is just the identity operator on the total Hilbert space, we can write the density matrix quite generally as

\begin{align}
\hat{\rho} &= \sum_{n',n''}^{N}\Pi_{n'}\hat{\rho}\Pi_{n''}. \nonumber
\end{align}

\noindent Then, using the block diagonal form of \eqref{eq:hat-A-block} for $\hat{A}$ in \eqref{eq:expt-trace}, we have

\begin{align}
\langle\hat{A}\rangle &= \mathrm{Tr}\left[\sum_{n',n''}^{N}\Pi_{n'}\hat{\rho}\Pi_{n''}\sum_{n}^{N}\Pi_{n}\hat{A}_{n}\Pi_{n}\right], \nonumber \\
&= \mathrm{Tr}\left[\sum_{n',n}^{N}\Pi_{n'}\hat{\rho}\Pi_{n}\hat{A}_{n}\Pi_{n}\right], \nonumber \\
&= \mathrm{Tr}\left[\sum_{n}^{N}\Pi_{n}\hat{\rho}\Pi_{n}\hat{A}_{n}\Pi_{n}\right]. \label{eq:expt-trace-ssr} 
\end{align}

\noindent In this last line, the reduction to a single summation over $n$ is due to the action of the trace operator singling out the diagonal elements. Moreover, the left and right-most projection operators $\Pi_{n}$ are redundant as their action will be accounted for by the combination of the inner $\Pi_{n}$ and the trace operator. It is, however useful to keep the left-most operator in place. Noting that $\Pi_{n}\hat{A}\Pi_{n} = \Pi_{n}\hat{A}_{n}\Pi_{n}$, we then obtain the expression

\begin{align}
\langle\hat{A}\rangle &= \mathrm{Tr}\left[\sum_{n}^{N}\Pi_{n}\hat{\rho}\Pi_{n}\hat{A}\right]. \nonumber
\end{align}

\noindent This makes it evident that the presence of a superselection rule is equivalent to using a general operator $\hat{A}$ with a density matrix of the form

\begin{align}
\hat{\rho} &\to \sum_{n}^{N}\Pi_{n}\hat{\rho}\Pi_{n}, \label{eq:density-matrix-effective}
\end{align}

\noindent which removes all the coherences between states lying in different subspaces $\H_{n}$. As we shall see in the Section~\ref{eq:effective}, there are circumstances under which the density matrix may well take this effective form.

\subsection{Physical observables}\label{eq:observables}
At this point, we return to the question asked earlier in Section~\ref{sec:definition}: \emph{when is a self-adjoint operator not a physical observable}? In attempting to answer this, we remind ourselves of the postulate of quantum mechanics that every physical observable is represented by a self-adjoint operator. In passing, we note that to many practising physicists the terms `Hermitian' and `self-adjoint' are used to mean the same thing, although in pure mathematics there is a subtle difference. Specifically, a Hermitian operator $\hat{A}$ must satisfy the Hermitian property $\langle\hat{A}\phi|\psi\rangle =\langle\phi|\hat{A}\psi\rangle$ for all $\ket{\phi}$ and $\ket{\psi}$ in the domain of $\hat{A}$. Whilst all self-adjoint operators are Hermitian, they must also have the property that the domains of $\hat{A}$ and $\hat{A}^{\dagger}$ are the same. We shall not enter into these subtleties here, but rather avoid the problem by sticking to the term `self-adjoint'. Instead, we shall consider the properties of a self-adjoint operator that make it suitable for modelling a physical observable.

An observable, by definition, must be something we can observe. In order to do this, we must interact with the system in some way to extract this information. This fixes the state of the system and hence the value of the observable we are trying to measure. Since physics is a quantifiable discipline, we are specifically concerned with quantities that we can compare. For this, we need to quantify observables using \emph{real} numbers.

Now a self-adjoint operator always has real eigenvalues, so this satisfies the real-valued requirement. Moreover, each eigenvector of the operator represents a possible state that the physical system might be found in when measuring the particular observable. Whenever we measure an observable quantity, we also set the system in the corresponding eigenstate. These then are essential features of observables that are well-modelled using self-adjoint operators.

Consider now how we might \emph{construct} a self-adjoint operator representing some observable of a system. Let us try asserting that some superposition of two orthonormal vectors $\ket{\psi_{1}}$ and $\ket{\psi_{2}}$ was an eigenvector of this operator, for example

\begin{align}
\ket{\xi} &= \mod{\alpha}\ket{\psi_{1}} + \mod{\beta}\rme^{\rmi\theta}\ket{\psi_{2}}, \nonumber
\end{align}

\noindent where we have explicitly written the amplitudes in terms of their relative phase and, as usual, we have $\sqmod{\alpha} + \sqmod{\beta} = 1$. Since the eigenvectors of a self-adjoint operator (with distinct eigenvalues) are orthogonal to each other, another eigenvector might be given by

\begin{align}
\ket{\xi'} &= \mod{\beta}\ket{\psi_{1}} - \mod{\alpha}\rme^{\rmi\theta}\ket{\psi_{2}}, \nonumber \\
&= \mod{\beta}\ket{\psi_{1}} + \mod{\alpha}\rme^{\rmi(\theta + \pi)}\ket{\psi_{2}}, \nonumber
\end{align}

\noindent using which we can easily verify $\iprod{\xi'}{\xi} = 0$. Note that in order to define this orthogonality condition, we \emph{require} the relative phase information. In this case, it is the phase shift of $\pi$ on the $\ket{\psi_{2}}$ that gives the factor of $-1$ that allows the orthogonality condition to be satisfied. 

Suppose, however, that a superselection rule existed separating $\ket{\psi_{1}}$ and $\ket{\psi_{2}}$. We saw in Section~\eqref{sec:superpositions} that this meant that the relative phase between the components was unobservable and hence unknowable. Without well-defined phase relations, no set of orthogonal vectors can be built from these separated vectors and therefore no valid observable can be constructed from them. Note that this is the case only because we have \emph{assumed} (presumably on physical grounds of some kind) that the Hilbert space must be decomposed into superselection sectors.

Let us now return to the self-adjoint operator introduced in Section~\ref{sec:definition}

\begin{align}
\hat{B} &= \oprod{\psi_{1}}{\psi_{2}} + \oprod{\psi_{2}}{\psi_{1}}. \nonumber
\end{align}

\noindent It was asserted there that this could not represent a physical observable because it violated the definition of the superselection rule \eqref{eq:superselection-rule}. However, it would seem to be fairly straight-forward to construct a basis set using $\ket{\psi_{1}}$ and $\ket{\psi_{2}}$ for which $\hat{B}$ is manifestly a self-adjoint operator. Namely

\begin{align}
\ket{\xi_{+}} &= \frac{1}{\sqrt{2}}\left(\ket{\psi_{1}} + \ket{\psi_{2}}\right) \nonumber
\end{align}

\noindent and

\begin{align}
\ket{\xi_{-}} &= \frac{1}{\sqrt{2}}\left(\ket{\psi_{1}} - \ket{\psi_{2}}\right), \nonumber
\end{align}

\noindent which have eigenvalues $\pm1$ when operated on by $\hat{B}$. The reason that $\hat{B}$ \emph{cannot} represent a physical observable is because the necessary phase relations between $\ket{\psi_{1}}$ and $\ket{\psi_{2}}$ required to construct the eigenvectors of $\hat{B}$ cannot be defined (specifically, the factor of $\pm1$ on $\ket{\psi_{2}}$). This in turn is because $\ket{\psi_{1}}$ and $\ket{\psi_{2}}$ are separated by a superselection rule. 

Given the circularity of this argument, the reader may be forgiven for considering superselection rules with some scepticism. However, in Section~\eqref{eq:effective} we shall consider cases in which such rules are not simply asserted but can \emph{effectively} arise as a natural consequence of the physics. 

Moreover, the work of Strocchi and Wightman~\cite{strocchi1974proof} discussed above offers a \emph{proof} that the charge operator commutes with all quasilocal variables, so that we then have a justification for the existence of a superselection rule based on physical arguments. 

\section{Effective superselection}\label{eq:effective}

\subsection{Absence of a shared reference frame}\label{sec:reference-frame}
It has been argued by Bartlett \emph{et al} that a superselection rule can effectively emerge due to the lack of a shared reference frame~\cite{bartlett2007reference}. For example, in quantum optics we typically model optical modes in the number representation via Fock states $\ket{n}$ (here, for simplicity, we shall just consider single-mode states). These are eigenstates of the occupation number operator $\hat{n}$. Applying the unitary operator $\hat{U}(\phi) = \exp(\rmi\phi\hat{n})$ to a Fock state, we would then get

\begin{align}
\hat{U}(\phi)\ket{n} &= \rme^{\rmi\phi n}\ket{n}, \nonumber
\end{align}

\noindent which adds a phase shift to the state.

Now, if this is not just taken to be an arbitrary global phase factor, this phase must be relative to some particular phase reference (say a high intensity laser beam) in an experimentalist's possession. Suppose, however that the experimentalist wished to share this state with a colleague who had no access to this reference. Any phase shift on the state might be taken to be the difference in phase between the two experimentalists frames of reference. But, without access to the reference phase, the second experimentalist has no way to determine what this phase difference is. The best that they can do is to average over all possible phases.

Let us analyse this situation by decomposing the Hilbert space of the system $\H$ into the form of  \eqref{eq:direct-sum} as the direct sum of subspaces $\H_{n}$ spanned by states for which $\hat{n}$ has eigenvector $n$. Applying the operator $\hat{U}(\phi) = \exp(\rmi\phi\hat{n})$ on any state $\ket{\psi_{n}}$ in $\H_{n}$, we would then have

\begin{align}
\hat{U}(\phi)\ket{\psi_{n}} &= \rme^{\rmi\phi n}\ket{\psi_{n}}. \nonumber
\end{align}

\noindent Defining $\Pi_{n}$ to be the projector onto $\H_{n}$ as previously, the action of $\hat{U}(\phi)$ on an arbitrary state $\ket{\psi}$ will then be

\begin{align}
\hat{U}(\phi)\ket{\psi} &= \sum_{n=0}^{\infty}\rme^{\rmi\phi n}\Pi_{n}\ket{\psi}. \nonumber
\end{align}

Now the action of $\hat{U}(\phi)$ models the advancement of the phase when the state is transferred between the experimentalists. In terms of the density matrix $\hat{\rho} = \oprod{\psi}{\psi}$, the averaging of the state over all possible phases may be formulated as

\begin{align}
\mathcal{U}\left(\hat{\rho}\right) &= \int_{0}^{2\pi} \hat{U}(\phi)\ket{\psi}\bra{\psi}\hat{U}^{\dagger}(\phi)~\frac{d\phi}{2\pi}, \nonumber \\
&= \sum_{n',n} \int_{0}^{2\pi} \rme^{\rmi\phi n'}\Pi_{n'}\hat{\rho}\rme^{-\rmi\phi n}\Pi_{n}~\frac{d\phi}{2\pi}, \nonumber \\
&= \sum_{n',n} \Pi_{n'}\hat{\rho}\Pi_{n} \int_{0}^{2\pi} \rme^{\rmi\phi (n'-n)}~\frac{d\phi}{2\pi}. \nonumber 
\end{align}

\noindent Now, since

\begin{align}
\int_{0}^{2\pi} \rme^{\rmi\phi (n'-n)}~\frac{d\phi}{2\pi} &= \delta_{n'n}, \nonumber 
\end{align}

\noindent we have

\begin{align} \nonumber \\
\mathcal{U}\left(\hat{\rho}\right) &= \sum_{n} \Pi_{n}\hat{\rho}\Pi_{n}. \nonumber
\end{align}

\noindent The action of this mapping is then to render the density matrix in the same form as \eqref{eq:density-matrix-effective}, the effective density matrix when a superselection rule pertains. This form then implies that coherent superpositions between states of different modal occupation number $n$ are not possible.


\subsection{Environment-induced superselection}
It was shown in Section~\ref{sec:superpositions} that the existence of a superselection rule was equivalent to the imposition of a block-diagonal density matrix in which coherences between different subspaces had been removed. This concept was used in Section~\ref{sec:reference-frame} where it was shown that such an effective density matrix can emerge in situations where two observers had no shared frame of reference.

In this section, we show how this effective form for the density matrix for a system interacting with a larger environment can emerge naturally as the result of phase information being lost to the environment through decoherence. This is known as \emph{environment-induced superselection} or `einselection'~\cite{zurek1982environment}.

In an open bipartite system, the total Hilbert space $\H$ may be taken to be the tensor product of a system space $\H_{\S}$ and that of its environment $\H_{\Env}$

\begin{align}
\H &= \H_{\S}\otimes\H_{\Env}. \nonumber
\end{align}

\noindent The total Hamiltonian of this joint system is then assumed to be composed of three parts, the self-Hamiltonians of the system $\hat{H}_{\S}$ and environment $\hat{H}_{\Env}$ plus an interaction Hamiltonian $\hat{H}_{I}$ acting on the total system

\begin{align}
\hat{H} &= \hat{H}_{\S} + \hat{H}_{\Env} + \hat{H}_{I}. \nonumber
\end{align}

For simplicity, here we shall assume that we are working in the \emph{quantum-measurement limit}~\cite{schlosshauer2007decoherence} in which the energies of the interaction Hamiltonian are much larger than either $\hat{H}_{\S}$ or $\hat{H}_{\Env}$ and consider only the action of $\hat{H}_{I}$. In this case, we can choose basis states that diagonalise $\hat{H}_{I}$ and write

\begin{align}
\hat{H}_{I} &= \sum_{ij} \hbar\w_{ij}\ket{\f_{i}}\bra{\f_{i}}\otimes\ket{\E_{j}}\bra{\E_{j}}, \nonumber
\end{align}

\noindent where the $\ket{\f_{i}}$ and $\ket{\E_{j}}$ are system and environmental states respectively. In the following, we shall take the notation $\ket{\f_{i}}\ket{\E_{j}}$ to imply the tensor product $\ket{\f_{i}}\otimes\ket{\E_{j}}$. A general state of the total system can then be written in terms of these states as

\begin{align}
\ket{\Psi} &= \sum_{ij} a_{ij}\ket{\f_{i}}\ket{\E_{j}}. \nonumber
\end{align}

\noindent Typically, we assume that the total system is initially in a tensor product of system and environmental states and put $a_{ij} = a_{i}b_{j}$. We then easily find the time evolution of this state by operating with $\exp(-\hat{H}_{I}t/\hbar)$, giving

\begin{align}
\ket{\Psi(t)} &= \sum_{ij} a_{i}b_{j}\rme^{-\rmi\w_{ij}t}\ket{\f_{i}}\ket{\E_{j}}. \label{eq:total-time-evolve}
\end{align}

\noindent Note that, in general, this will be an \emph{entangled} state since it is no longer possible to decompose it into a tensor product of system and environmental states. Note that this is due to the \emph{phase} factors $\rme^{-\rmi\w_{ij}t}$ coupling the product terms.

It will be convenient to re-express \eqref{eq:total-time-evolve} as

\begin{align}
\ket{\Psi(t)} &= \sum_{i} a_{i}\ket{\f_{i}}\ket{\R_{i}(t)}, \nonumber
\end{align}

\noindent where 

\begin{align}
\ket{\R_{i}(t)} &\equiv \sum_{j} b_{j}\rme^{-\rmi\w_{ij}t}\ket{\E_{j}} \nonumber
\end{align}

\noindent is the \emph{relative state}~\cite{everett1956theory} associated with the system state $\ket{\f_{i}}$. Of particular interest will be the inner products of relative states

\begin{align}
\iprod{\R_{i'}(t)}{\R_{i}(t)} &= \sum_{j} \sqmod{b_{j}}\rme^{\rmi(\w_{i'j} - \w_{ij})t}. \label{eq:R-inner-prod}
\end{align}

\noindent If the $\w_{ij}$ are all distinct, then for $i'\ne i$ this expression \emph{may} (but not necessarily) give a function that tends to zero as $t$ tends to infinity. For example, under certain model conditions, \eqref{eq:R-inner-prod} might be made to take the form of a sinc function that decays on a characteristic time scale~\cite{vaughan2020concept}. The effect of this would be to suppress coupling between system states $\ket{\phi_{i}}$ and environmental relative states $\ket{\R_{i'}}$ where $i' \ne i$ and hence leads to a strong one-to-one correlation between the system and relative states where $i = i'$. Zurek therefore refers to $\iprod{\R_{i'}(t)}{\R_{i}(t)}$ as the \emph{correlation amplitude}~\cite{zurek1982environment}.

The process of decoherence is modelled by tracing out the environmental degrees of freedom from the density matrix of the total system. This is achieved by performing the partial trace of just the environmental states, leaving the \emph{reduced density matrix} of the system~\cite{schlosshauer2007decoherence}

\begin{align}
\hat{\rho}_{\S} &= \mathrm{Tr}_{\Env}\left[\hat{\rho}\right]. \label{eq:reduced-rho}
\end{align}

\noindent This gives us a density matrix on just the system Hilbert space $\H_{\S}$ but that has incorporated the interaction with the environment. 

Inserting $\hat{\rho} = \ket{\Psi(t)}\bra{\Psi(t)}$ into \eqref{eq:reduced-rho}, we find

\begin{align}
\hat{\rho}_{\S}(t) &= \sum_{i'i} a_{i}a_{i'}^{*}\iprod{\R_{i'}(t)}{\R_{i}(t)}\ket{\phi_{i}}\bra{\phi_{i'}}. \label{eq:reduced-rho-S}
\end{align}

\noindent At $t=0$, all the relative state inner products will be equal to unity and \eqref{eq:reduced-rho-S} reduces to the pure-state density matrix for an isolated system. However, as the system evolves the relative state inner products may drive the off-diagonal matrix elements to zero, resulting in a mixed-state density matrix. In the case where all the $\w_{ij}$ are distinct, $\hat{\rho}_{\S}(t)$ may then evolve to take the diagonal form

\begin{align}
\lim_{t\to\infty} \hat{\rho}_{\S}(t) &= \sum_{i} \sqmod{a_{i}}\ket{\phi_{i}}\bra{\phi_{i}}, \label{eq:reduced-rho-S-limit}
\end{align}

\noindent which is the density matrix for a classical ensemble. Note that the interaction Hamiltonian $\hat{H}_{I}$ is picking out the system states $\ket{\phi_{i}}$ that become correlated with the relative states of the environment. Hence, observables constructed from these states will be those measurable under the system-environment interaction. These states have been called \emph{pointer states} by Zurek~\cite{zurek1981pointer}.

In the more general case, it may be that some of the energies coupling system and environmental states in the interaction Hamiltonian are degenerate. In this case we might rewrite $\hat{H}_{I}$ as

\begin{align}
\hat{H}_{I} &= \sum_{n}\sum_{i_{n}j} \hbar\w_{i_{n}j}\ket{\f_{i_{n}}}\bra{\f_{i_{n}}}\otimes\ket{\E_{j}}\bra{\E_{j}}, \nonumber
\end{align}

\noindent where, for a given $n$ and $j$ all the energies $\hbar\w_{i_{n}j}$ are equal. We may then view each set of system states $\ket{\f_{i_{n}}}$ as belonging to a subspace $\H_{n}$ of the system Hilbert space and decomposing this into a direct sum as given by \eqref{eq:direct-sum}.

Denoting the value of these common energies by $\hbar\w_{nj}$, the expression for the coupling amplitude then becomes modified to

\begin{align}
\iprod{\R_{i_{m}'}(t)}{\R_{i_{n}}(t)} &= \sum_{j} \sqmod{b_{j}}\rme^{\rmi(\w_{mj} - \w_{nj})t}, \nonumber
\end{align}

\noindent where $i_{n}$ labels the states of $\H_{n}$. In the cases where the correlation amplitude tends to zero as $t$ tends to infinity, we will now have the evolution

\begin{align}
\lim_{t\to\infty}\iprod{\R_{i_{m}'}(t)}{\R_{i_{n}}(t)} &= \delta_{mn}. \label{eq:deltann}
\end{align}

\noindent Note that for states in the same subspace, we will have (in this simple model)

\begin{align}
\iprod{\R_{i_{n}'}(t)}{\R_{i_{n}}(t)} &= \sum_{j} \sqmod{b_{j}}\rme^{\rmi(\w_{nj} - \w_{nj})t} = \sum_{j} \sqmod{b_{j}} = 1. \nonumber
\end{align}

\noindent Hence, within a given subspace, there will be no individual coupling of system states to relative states. Rather, as we shall see, each state of the subspace will be coupled to the \emph{same} environmental relative state. For this reason, we can simplify the notation and simply denote the $i_{n}$th relative state by $\ket{\R_{n}(t)}$.

Employing the projectors $\Pi_{n}$ on the $n$th system subspace, the reduced density matrix of \eqref{eq:reduced-rho-S} is equivalent to

\begin{align}
\hat{\rho}_{\S}(t) &= \sum_{mn}\sum_{i_{m}'i_{n}} \Pi_{n}a_{i_{n}}a_{i_{m}'}^{*}\iprod{\R_{m}(t)}{\R_{n}(t)}\ket{\phi_{i_{n}}}\bra{\phi_{i_{m}'}}\Pi_{m}. \nonumber
\end{align}

\noindent Now, however, using \eqref{eq:deltann} the evolution is 

\begin{align}
\lim_{t\to\infty}\hat{\rho}_{\S}(t) &= \sum_{n}\sum_{i_{n}'i_{n}} \Pi_{n}\ket{\phi_{i_{n}}}a_{i_{n}}a_{i_{n}'}^{*}\bra{\phi_{i_{n}'}}\Pi_{n}, \nonumber \\
&= \sum_{n}\Pi_{n}\hat{\rho}_{\S}(0)\Pi_{n}, \nonumber
\end{align}

\noindent which has the same form as the effective density matrix of \eqref{eq:density-matrix-effective} in the presence of a superselection rule. Note that $\hat{\rho}_{\S}(0)$ is a pure-state density matrix, so that within a given subspace $\H_{n}$ coherent superpositions are still possible but that coherences between different subspaces have now been removed. This then demonstrates how an \emph{effective} superselection rule may emerge for a system as a consequence of decoherence with its environment.

In contrast to the case of charge conservation described above, the superselection rule emerging here is, in general, only approximate. However, for the simple model described in this section, it will become exact when the relative states associated with each subspace $\H_{n}$ become exactly orthogonal to each other. With only the interaction Hamiltonian acting on the system, each system state of $\H_{n}$ shares the same energy expectation value, which manifests somewhat like an `effective eigenvalue'. 

To see this, consider the tensor product component of the total state

\begin{align}
\ket{\Psi_{i_{n}}(t)} &= \ket{\f_{i_{n}}}\ket{\R_{n}(t)}, \label{eq:Psi-in}
\end{align} 

\noindent where $\ket{\f_{i_{n}}} \in \H_{n}$ and

\begin{align}
\ket{\R_{n}(t)} &= \sum_{j} b_{j}\rme^{-\rmi\omega_{nj}t}\ket{\E_{j}} \nonumber
\end{align} 

\noindent (recall $\omega_{nj} = \omega_{i_{n}j}$ for all $i_{n}$). Acting on \eqref{eq:Psi-in} with the interaction Hamiltonian $\hat{H}_{I}$, we have

\begin{align}
\hat{H}_{I}\ket{\Psi_{i_{n}}(t)} &= \ket{\f_{i_{n}}}\sum_{j} \hbar\omega_{nj}b_{j}\rme^{-\rmi\omega_{nj}t}\ket{\E_{j}}.
\end{align} 

\noindent Projecting on the left with $\ket{\R_{n}(t)}$, we get

\begin{align}
\bra{\R_{n}(t)}\hat{H}_{I}\ket{\Psi_{i_{n}}(t)} &= \expt{\hbar\w_{n}}\ket{\f_{i_{n}}}, \nonumber
\end{align} 

\noindent where

\begin{align}
\expt{\hbar\w_{n}} &= \sum_{j}\sqmod{b_{j}}\hbar\w_{nj} = \iprodM{\Psi_{i_{n}}}{\hat{H}_{I}}{\Psi_{i_{n}}}, \nonumber
\end{align}

\noindent which is the average of $\hbar\w_{nj}$ taken over all the environmental states.

\section{Physical processes}
\subsection{Unitary evolution}

Mathematically, one may always transform a normalised vector of a Hilbert space into another via a unitary transformation. This is accomplished via a unitary operator $\hat{U}$, the inverse of which, by definition, is equal to its conjugate transpose

\begin{align}
\hat{U}^{\dagger} &= \hat{U}^{-1}. \label{eq:unitary-op-def}
\end{align}

\noindent It is then straight-forward to see that a unitary preserves the normalisation of any vector it transforms, since for any vector $\ket{\Phi} = \hat{U}\ket{\Psi}$, where $\ket{\Psi}$ is normalised, we will have $\iprod{\Phi}{\Phi} = \iprodM{\Psi}{\hat{U}^{\dagger}\hat{U}}{\Psi} = \iprod{\Psi}{\Psi} = 1$. Unitary operators are essential for the description of physical processes in quantum mechanics, since by preserving the normalisation of a vector, they preserve the physical information encoded into it.

Of particular significance is the fact that any unitary operator $\hat{U}$ may be related to an associated self-adjoint operator $\hat{H}$ via 

\begin{align}
\hat{U}(t) &= \rme^{\rmi\hat{H}t}, \label{eq:unitary-hermitian}
\end{align}

\noindent where $t$ is a real, scalar parameter. The canonical example is, of course, time evolution, where $\hat{H}$ is the Hamiltonian of the system and $t$ is taken to be time. Now, we have already seen that in the presence of a superselection rule, not \emph{every} self-adjoint operator can be interpreted as physical observable. This begs the related question, can we associate \emph{every} unitary operator with a \emph{physical process}?

Of course, before we can answer this question, we must specify exactly what we mean by a `physical process'. To make headway on this, we shall \emph{define} a physical process to be a transformation that can be modelled by a unitary operator of the form of \eqref{eq:unitary-hermitian} where $\hat{H}$ \emph{is a physical observable}. 

We should also take care to distinguish between `active' unitary transformations, as opposed to `passive' transformations, such as a change of basis. An `active' transformation, such as time evolution, implies that something has actually changed about the state being represented, whereas a change of basis (which would be applied to both states \emph{and} operators) implies no such change.

\subsection{No evolution between separated states}
Let us consider a unitary operator $\hat{U}$ defined in terms of a physical observable $\hat{A}$ via

\begin{align}
\hat{U}(t) &= \rme^{\rmi\hat{A}t}, \label{eq:physical-process}
\end{align}

\noindent where $t$ is some real, scalar parameter. According to our definition above, the action of this operator on a vector of the Hilbert space will then constitute a physical process.

Expanding $\hat{U}(t)$ as a power series, we will have

\begin{align}
\hat{U}(t) &= \hat{I} + \rmi t\hat{A} + \frac{(\rmi t)^{2}}{2} \hat{A}^{2} + \ldots \nonumber
\end{align}

\noindent or, in terms of the eigenvalues $A_{i}$ and eigenvectors $\ket{\phi_{i}}$ of $\hat{A}$,

\begin{align}
\hat{U}(t) &= \sum_{i}\left[1 + \rmi tA_{i} + \frac{(\rmi tA_{i})^{2}}{2} + \ldots\right]\ket{\phi_{i}}\bra{\phi_{i}}. \nonumber
\end{align}

Now, if the Hilbert space is decomposed into superselection sections $\H_{n}$, since $\hat{A}$ is a physical observable, we should re-write this as

\begin{align}
\hat{U}(t) &= \sum_{n}\sum_{i_{n}}\left[1 + \rmi tA_{i_{n}} + \frac{(\rmi tA_{i_{n}})^{2}}{2} + \ldots\right]\ket{\phi_{i_{n}}}\bra{\phi_{i_{n}}}, \nonumber
\end{align}

\noindent where $\ket{\phi_{i_{n}}} \in \H_{n}$. Acting on a state $\ket{\psi_{n}} \in \H_{n}$, we then have

\begin{align}
\hat{U}(t)\ket{\psi_{n}} &= \sum_{i_{n}}\left[1 + \rmi tA_{i_{n}} + \frac{(\rmi tA_{i_{n}})^{2}}{2} + \ldots\right]\ket{\phi_{i_{n}}}\iprod{\phi_{i_{n}}}{\psi_{n}}, \nonumber
\end{align}

\noindent where the terms associated with other superselection sectors have dropped out. Therefore, projecting a state $\ket{\psi_{n'}}$ onto this where $n' \ne n$ (and hence $\ket{\psi_{n'}}$ belongs to a different sector), we will have

\begin{align}
\iprodM{\psi_{n'}}{\hat{U}(t)}{\psi_{n}} &= 0. \nonumber
\end{align}

\noindent In other words, \emph{no physical process (as defined above) can evolve a state from one superselection sector to another}.

\section{Is there an energy superselection rule?}
\subsection{Plausibility arguments}
The arguments for a charge superselection rule as laid out in Section~\ref{sec:charge} are highly compelling when supported by Strocchi and Wightman's formal proof~\cite{strocchi1974proof}, and it is widely held that such a rule exists. In 1983, Page and Wootters went beyond this to suggest that an analogous rule might exist for energy, with deeply profound implications~\cite{page1983evolution}. They argued that since the gravitational field is just as long-range as the Coulomb field, an analogous argument to that of Strocchi and Wightman might also hold for quasilocal operators commuting with the total Hamiltonian of the system.  

Intuitively, we might also argue for this on the grounds of energy conservation, which is generally accepted as a universal physical principle. Whilst a completely isolated physical system might be decomposed into interacting subsystems, each with their own self Hamiltonians and coupled via interaction Hamiltonians, the total energy of the \emph{total} system ought to be conserved. This might be interpreted as saying that no physical observable ought to be able to couple energy eigenstates of the total system Hamiltonian with different eigenvalues. Hence, there would be a superselection rule due to total energy.

\subsection{The unobservability of coordinate time}
If such a superselection rule based on total energy did exist, this would imply that all physical observables will commute with the total Hamiltonian of the system. This, in turn, implies that every physical observable must then be a constant of the motion, which begs the question of how time evolution can be observed at all. Page and Wooters refer to such a time frame for the total system as \emph{coordinate time}, a term adopted from relativity theory. Under an energy superselection rule, such coordinate time must be unobservable.

Page and Wooters then go on to describe how one may still describe some kind of temporal evolution by taking the pointer states of a subsystem to provide a form of `clock time', relative to which other systems may be seen to evolve. The development of these ideas is truly fascinating but unfortunately takes us beyond the scope of this current article. Nonetheless, given the profound implications of an energy superselection rule, the idea deserves very serious consideration. 

\section{Conclusions}
We have seen that the possibility of a superselection rule on a Hilbert space imposes certain restraints on what we can consider to be a physical observable. Moreover, since coherent superpositions of states from separated superselection sectors would no longer possible, we must also face the possibility that not every unit vector of a Hilbert space can represent a unique physical state. If we then define a physical process in terms of a unitary transformation that can be expressed in terms of a physical observable, we would further need to accept that not every active unitary transformation that we can define mathematically corresponds to a valid physical process. 

Hence, superselection rules (if they exist) will impose certain limitations on how we use the \emph{mathematics} of Hilbert spaces to model \emph{physics}. The arguments for the superselection rules themselves emerge (it would seem), not from the mathematics \emph{per se}, but from the physical arguments for charge or possibly energy conservation. There is also a strong case for suggesting that \emph{effective} superselection rules may emerge out of decoherence or the lack of a shared reference frame.

\subsection{Summary}
Given the significance of these consequences for our fundamental understanding of what constitutes a physical state, observable or process, it is important for physicists to have a solid understanding of concept of superselection. Here, we conclude by giving a brief summary of the main points discussed in this paper.

\begin{itemize}

\item [(1)] \emph{If for \emph{\textbf{any}} physically realisable observable $\hat{A}$, we have}
\begin{align}
\iprodM{\psi_{1}}{\hat{A}}{\psi_{2}} &= 0, \nonumber
\end{align}
 \emph{then $\ket{\psi_{1}}$ and $\ket{\psi_{2}}$ are said to be separated by a \emph{\textbf{superselection rule}}}.
 
  \item [(2)] \emph{The existence of a superselection rule on a Hilbert space $\H$ implies that not every self-adjoint operator that can be mathematically defined on $\H$ represents a physical observable.}

 \item [(3)] \emph{If a superselection rule exists on a Hilbert space $\H$, it can be decomposed into a direct sum}
 
 \begin{align}
\H &= \bigoplus_{n} \H_{n} \nonumber
\end{align}

\emph{where the subspaces $\H_{n}$ are referred to as \emph{\textbf{superselection sectors}}, such that no physical observable can couple states from different $\H_{n}$.}

\item [(4)] \emph{Coherent superpositions of states from different superselection sectors are not possible.}

\item [(5)] \emph{A superselection rule exists on a Hilbert space $\H$ between the sets of eigenvectors of an operator $\hat{A}$ having the different eigenvalues $A_{n}$ and  $A_{n'}$ if and only if $\hat{A}$ commutes with \emph{\textbf{all}} other physical observables}.

\item [(6)] \emph{Mathematically, a physical observable for which a superselection rule exists can be modelled by a general operator in conjunction with an effective density matrix of the form}

\begin{align}
\hat{\rho}_{eff} &= \sum_{n}\Pi_{n}\hat{\rho}\Pi_{n}, \nonumber
\end{align}

\emph{where $\hat{\rho}$ is the unmodified density matrix and the $\Pi_{n}$ are the projection operators onto the superselection sectors $\H_{n}$.}

\item [(7)] \emph{Effective superselection rules may emerge dynamically via, for example, a lack of a shared reference frame or through decoherence of a system with its environment. In these cases, the superselection rule is induced by the emergence of an effective density matrix $\hat{\rho}_{eff}$ of the form given above}.

\item [(8)] \emph{Defining a} \textbf{physical process} \emph{as an active unitary transformation of the form}

\begin{align}
\hat{U}(t) &= \rme^{\rmi\hat{A}t}, \nonumber
\end{align}

\emph{where $\hat{A}$ is a physical observable and $t$ is a real scalar, it can be shown that no physical process can evolve a state from one superselection sector to another.}

\end{itemize}

\bibliographystyle{unsrt}
\bibliography{refs}

\end{document}